\documentclass[runningheads]{llncs}
\usepackage{graphicx}
\usepackage{amsmath}
\usepackage{tikz}
\usepackage{graphicx}
\usepackage{caption}
\usepackage{amsfonts}
\usepackage{makecell}
\usepackage{subcaption}

\usepackage{amsthm}
\usepackage[english]{babel}
\usepackage{booktabs}
\usepackage{cite}

\begin{document}
\title{IACN: Influence-aware and Attention-based Co-evolutionary Network for Recommendation}
%
%
\author{Shalini Pandey \and
George Karypis \and
Jaideep Srivasatava}
%

\institute{Department of Computer Science and Engineering\\University of Minnesota\\ Twin Cities, Minnesota, USA\\
\email{\{pande103,karypis, srivasta\}@umn.edu}}

\maketitle              
\begin{abstract}
Recommending relevant items to users is a crucial task on online communities such as Reddit and Twitter. For recommendation system, representation learning presents a powerful technique that learns embeddings to represent user behaviors and capture item properties. However, learning embeddings on online communities is a challenging task because the user interest keep evolving. This evolution can be captured from 1)  interaction between user and item, 2) influence from other users in the community. The existing dynamic embedding models only consider either of the factors to update user embeddings. However, at a given time, user interest evolves due to a combination of the two factors. To this end, we propose Influence-aware and  Attention-based Co-evolutionary Network (IACN). Essentially, IACN consists of two key components: interaction modeling and influence modeling layer. The interaction modeling layer is responsible for updating the embedding of a user and an item when the user interacts with the item. The influence modeling layer captures the temporal excitation caused by interactions of other users. To integrate the signals obtained from the two layers, we design a novel fusion layer that effectively combines interaction-based and influence-based embeddings to predict final user embedding.  Our model outperforms the existing state-of-the-art models from various domains.

\end{abstract}
\section{Introduction}
Online communities such as Facebook, Twitter, and Reddit are a crucial part of today's online world. Recommendation of relevant information is essential for these platforms to improve users' experience and maintain their long-term engagement. However, the recommendation task involves various challenges. First, when a user interacts with an item both the user and item features are updated. Second, since users share information on online communities, they are likely to influence each other. For instance, when a user posts a comment on a thread in Reddit, she influences other users to post comments on the thread. Further, the degree of influence of interaction is dependent on the relation between the users and the time elapsed since the interaction. Naturally, as more time elapses the degree of influence reduces ~\cite{iwata2013discovering}.
Third, even when a user does not take any action, her interest keeps evolving ~\cite{kumar2019predicting}. It is important to determine user interest at any query time which can be predicted by the information from both her interactions and the influence of other users. \par

\begin{figure*}[t]
  \begin{subfigure}[b]{0.33\textwidth}
    \includegraphics[width=0.9\linewidth]{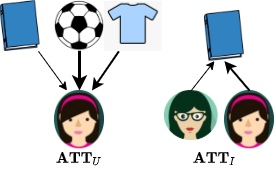}
    \caption{Interaction modeling layer}
    \label{intro1}
  \end{subfigure}\hfill
  \begin{subfigure}[b]{0.32\textwidth}
    \includegraphics[width=0.9\linewidth]{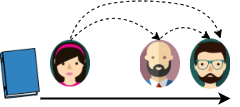}
    \caption{Influence modeling layer}
    \label{intro2}
  \end{subfigure}\hfill
  \begin{subfigure}[b]{0.32\textwidth}
    \includegraphics[width=0.9\linewidth]{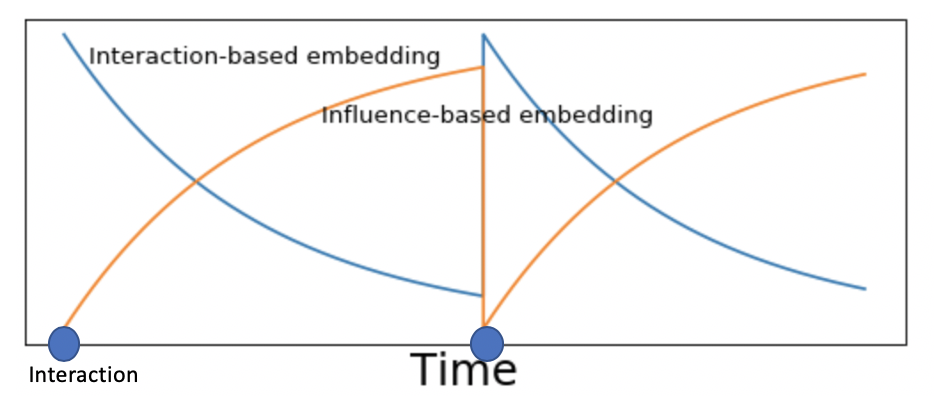}
    \caption{Contribution of interaction model and influence model towards user embedding with time}
    \label{intro3}
    \end{subfigure}
     \caption{A simplified diagram showing the main components of IACN. }
\end{figure*}

The existing research has individually addressed either of the above-mentioned challenges.
To capture the co-evolving nature of item properties and user interests, co-evolutionary models were proposed in ~\cite{dai2016deep,lu2018coevolutionary}. These models update user embedding only when the user interacts with an item even when user interest keeps evolving with time.  To capture the change in user's interest over time, JODIE ~\cite{kumar2019predicting} employs a projection operation that takes the user embeddings and time elapsed since the last user interaction to predict the user's embedding at any query time. Our model, instead, relies on the idea that users' interests at any future time can be predicted by the influence of other users on the user. To this front, several works have been done to augment the information from the influence of other users for predicting user interest ~\cite{fan2019graph, song2019session}. However, in these methods the influence of other users is static ~\cite{fan2019graph} or context-dependent ~\cite{song2019session}. Furthermore, they generate static embeddings of items; thus, ignoring their evolving properties.  \par


In this paper, we exploit both interaction and influence information for predicting temporal embeddings of users and items given the interaction sequence. The motivation is that when a user interacts with an item her interest at that time can be determined from the interaction features. However, as time elapses, the interest of the user drifts and tends to be more driven by the influence of other users. The key components in the IACN model are:\\
\textbf{Interaction modeling layer:} The interaction modeling layer is responsible for updating the embedding of corresponding users and items when they interact with each other.  We leverage the attention mechanism to identify which interactions are important for determining the updated embedding of entities (users and items) involved in the interaction.
As shown in Figure ~\ref{intro1}, when a user interacts with an item, ATT\textsubscript{U}  updates the embedding of the user by adaptively assigning weights to its previous interactions. Similarly, ATT\textsubscript{I} updates the embedding of the item based on its past interaction.
\\
\textbf{Influence modeling layer: }  We design a "relation revealing" attention-based operation to capture the relation between users and then update the embedding of a user when any user who influences the user interacts with an item.  As shown in Figure ~\ref{intro2}, when a user interacts with an item, it triggers a drift of interest of other users towards the item.
\\
\textbf{Fusion layer:} To learn future embedding of a user, we design a novel fusion layer that integrates the embedding from interaction and influence modeling layer. When an interaction occurs, the user embedding is determined solely by the interaction modeling layer because user interaction reveals the user's current interest ~\cite{kumar2019predicting}. As time progresses user embedding drifts further apart from the interaction embeddings. As shown in Figure ~\ref{intro3},  the future user embedding is computed by additively combining the influence-based embedding and the interaction-based embedding where the contribution of the interaction model decays while that of the influence model increases with time. \par

To recommend the next item which the user will interact with, IACN predicts an embedding for the next item and uses Locality Sensitive Hashing ~\cite{gionis1999similarity} to find the item whose embedding is most similar to the predicted item embedding. Extensive experimentation on real-world datasets shows that IACN outperforms six state-of-the-art models for the next item prediction task.  Also, we conduct a comprehensive ablation study to show the effect of key components.
Summary of our paper major contributions are:

\begin{itemize}
\item We study the contribution of both the interaction model and the influence model in predicting embeddings for the recommendation.
\item We design a co-evolutionary network using two attention layers to update the embeddings of users and items.  The attention layers help in improving the performance of our model along with providing  insight into different user behaviors.

\item We introduce a novel method to model the influence of other users on a user and integrate it with the interaction model to obtain the user embedding at query time
\item We conduct experimentation on the real-world dataset and demonstrate the superiority of our model over state-of-the-art baselines over various domains.
\end{itemize}
\vspace{-4mm}
\section{Related Work}
\vspace{-4mm}
\textbf{Dynamic co-evolutionary models.}
Joint modeling of users and items has been explored in recommendation systems. Models that concurrently learn both user and item embeddings have been developed in work such as ~\cite{wu2017recurrent, kumar2019predicting,dai2016deep}.
Methods such as ~\cite{wu2017recurrent, dai2016deep} use Recurrent Neural Nets (RNNs) to model the evolving features of items and users. They jointly learn representations of users and items with the idea that user and item embeddings influence each other whenever they interact. However, a user's interest changes with time even when he/she is not interacting with any item. JODIE ~\cite{kumar2019predicting} attempts to take into account the dynamic interest of users and update users' embedding by scaling their past embedding with a time context vector. Compared to other co-evolutionary models, we attempt to utilize self-attention mechanism to generate new embeddings since self-attention mechanism are more interpretable and have shown better performance at sequence modeling task ~\cite{vaswani2017attention} compared to the RNN-based method. In addition, we take into account the social influence of interaction  by a user in evolving the interest of other users. This helps in predicting future embedding trajectory of users. \\
\textbf{Recurrent Neural Networks based models.}
Several  models employ recurrent neural networks (RNNs) or their variants (LSTMs and GRUs) to build recommender systems ~\cite{hidasi2015session, wu2017recurrent}.
Models such as LSTM ~\cite{hidasi2015session} consider item embedding to be static which does not change with time. Unlike these models, RRN ~\cite{wu2017recurrent} uses two RNN layers to generate dynamic user and item embeddings from ratings data. However RRN only considers static embedding of items and users as inputs.
IACN, on the other hand, assigns each user and item both dynamic and static embedding. Also, IACN  considers both the update of embedding caused by interactions between user and item and influence  between users. \\
\textbf{Social Recommendation model.}
User's interest is affected by the neighboring members in social community.
 Some work ~\cite{song2019session, chen2019social} in  social recommendation need information of social network structures to
 predict the social influence. However, such structural information is not always available. Even if the social structure exist, the degree of influence of a user on another is rarely explicitly declared. As a result, we can rely only on estimating influences and relations among users from their activities.
These interactions  are implicit in the sense that users interact with one another by expressing their preferences on shared items ~\cite{xia2009ballot}. Modeling influence between users in the absence of knowledge of topology structure has been done in ~\cite{iwata2013discovering}.
To incorporate dynamic user interaction,  ~\cite{iwata2013discovering}  uses a Poisson process for modeling the influence of one user over the other where the influence has an exponential time decaying factor. They model the social
influence as a combination of features extracted from the users'
 behavior and features associated with their interactions with items. Finally, GraphRec ~\cite{fan2019graph} is a state-of-the-art model that utilizes both the interaction network and social network to predict user interest. Our model, however, differs from GraphRec as we consider the temporal dynamics involved in predicting embeddings of users by taking interactions of users and the influence from social network.\\
 \textbf{Temporal network embedding.}
Several models have recently been developed that generate embeddings for the nodes
in temporal networks where nodes are continuously added and links between nodes keep changing. Such models have been employed for recommendation task as well constituting the time-aware recommendation models where the nodes comprise of users and items.
Continuous-Time Dynamic Network Embeddings (CTDNE) \cite{nguyen2018continuous} generates embeddings using temporally evolving random walks, but it generates one final static embedding of the nodes.
HTNE ~\cite{zuo2018embedding} is the state-of-the-art  model for generating the embeddings of nodes in temporal networks. It utilizes Hawkes Process and attention mechanism to predict future interaction. They model the likelihood of interaction between two nodes using Hawkes process to capture the influence of their historical neighbors on each other. IACN, also models the temporally evolving network and generates embeddings to represent the feature of nodes that evolves with time. However, IACN considers both the interaction and social network for predicting embeddings of entities.
\vspace{-4mm}
\section{Notations, Definitions, and Preliminaries}
\vspace{-4mm}
\textbf{Notations.}
Given $m$ users and $n$ items, we denote the temporal list of $N$ observed interactions as $\mathcal{O} = \{o_j = (u_j , i_j , t_j , \boldsymbol{q}_j ) \forall j\in N \}$, where $u_j \in \{1, \ldots, m\}, i_j \in \{1, \ldots, n\}, t_j \in \mathbb{R}^+$ and $\boldsymbol{q}_j \in \mathbb{R}^F$ represent the interaction features. For simplicity, we define $\mathcal{O}^u = \{o_j^u=(i_j,t_j,\boldsymbol{q}_j)\}$ as the ordered listed of all interactions related to user $u$, and  $\mathcal{O}^i = \{ o_j^i = (u_j, t_j, \boldsymbol{q}_j)\}$ as the ordered list of all interactions related to item $i$.




In addition, users are influenced by other users in the social network.
When we arrange different user's interaction with items as a sequence according to ascending time, we can find which users influence other users to interact with the item.  These users form local user neighborhood for the user in consideration. As the user interacts with more items, its neighborhood keeps evolving. We can formally define the local user neighborhood of a user as follows:
\vspace{-1mm}
\begin{definition}{\textbf{Local user neighborhood}}
Given a temporal interaction network $G = <V, E; \mathcal{O}>$ representing the observed user-item interactions, the local user neighborhood, $\mathcal{N}_u(t)$ of a user $u$  are all those users $v \in U$ which are associated with at least one item before $u$  interacted with it.  Mathematically, when an interaction $o_j =(u_j,i_j,t_j,\boldsymbol{q}_j)$ is observed the local user neighborhood is updated as, $\mathcal{N}_u(t_j) = \mathcal{N}_u(t_j^-) \cup \mathcal{U}^i(t_j^-)$, where $ \mathcal{U}^i(t_j^-)$ is the set of user who interacted with item $i$ before time $t_j$ and $\mathcal{N}_u(t_j^-) $ is the local neighborhood of user $u$ right before time $t_j$.
\end{definition}
 \theoremstyle{definition}

\vspace{-6mm}
\section{Proposed Method}
\begin{figure*}[t]
       \includegraphics[width=\textwidth]{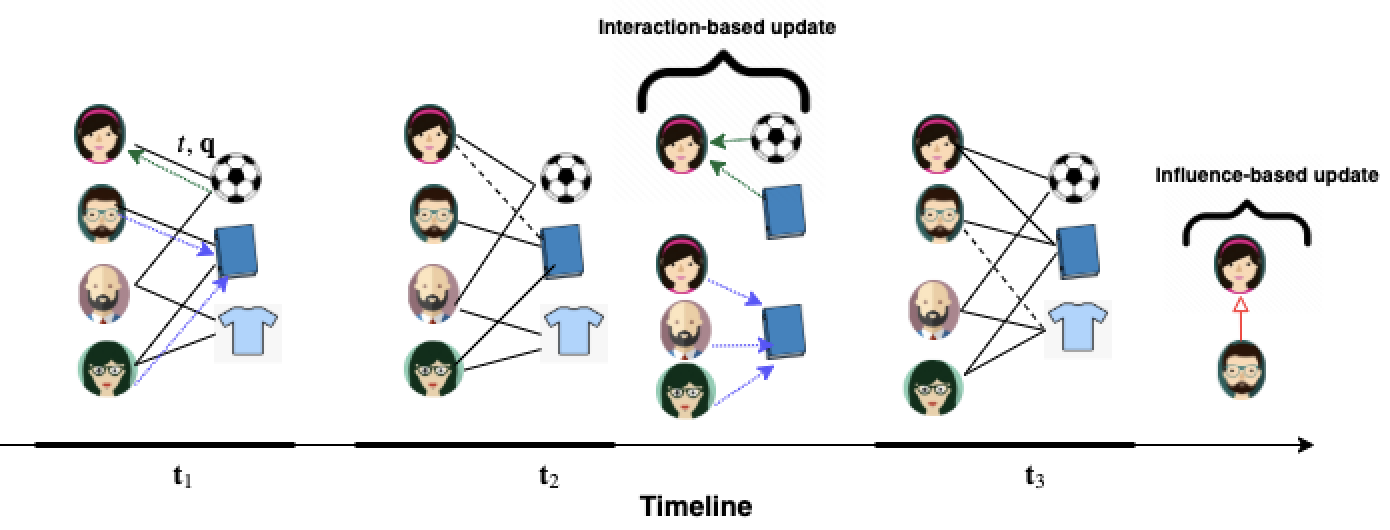}
     \caption{Model Illustration: Temporal Interaction Network at different timestamps. Green dashed arrows and blue dashed arrows indicate the attention of items in computing user embedding and users in computing item embedding, respectively.  Black dashed line refers to a new interaction and red arrow indicates the update in girl's embedding caused by an interaction by a user in her neighborhood at time $t_3$.
 }
     \label{architecture}

   \end{figure*}

\vspace{-2mm}
In this section, we introduce our proposed model, Influence-aware and Attention-based Co-evolutionary Network (IACN),  see Figure ~\ref{architecture} for a visual depiction of the architecture.
 To model the evolution of embeddings, IACN employs two layers to model the embedding update caused by interaction and influence and one layer to integrate the embeddings obtained from the modeling layers. \\
\textbf{ Interaction modeling layer:}  The interaction modeling layer consists of two attention functions, one to update user embedding (ATT\textsubscript{U}) and the other to update item embedding (ATT\textsubscript{I}). When an interaction $o = (u,i,t,\textbf{q})$ is observed, ATT\textsubscript{U} (resp., ATT\textsubscript{I})  updates the embedding of $u$ (resp., $i$).  \\
\textbf{ Influence modeling layer:}  The influence modeling layer uses the local user neighborhood for predicting user future interest. When one of local user neighbor interacts with an item, it triggers the user to interact with the item. This results in update of the user embedding. The degree of influence is determined by the relation between the users and the time elapsed since the interaction. \\
\textbf{Fusion layer:} This layer  predicts the user embedding at a future time since its last interaction by integrating both its interaction-based embedding and influence-based embedding.
\vspace{-5mm}

\subsection{Model Details}
\vspace{-2mm}

We will now describe each layer in IACN in detail.\\
\textbf{Embedding layer.}
We assign each user and item two embeddings: a static and a dynamic embedding. The static embedding encodes
 the long-term stationary properties while the dynamic embedding encodes the dynamic properties. This decision is made by following the setting in ~\cite{kumar2019predicting} such that
static embeddings, for a user, $u$,
$\boldsymbol{\bar{u}}  \in  \mathbb{R}^m$ and item $i$,
$\bar{\boldsymbol{i}} \in \mathbb{R}^n$ represent the long-term properties of the entities.  While dynamic embeddings $\boldsymbol{u}(t) \in \mathbb{R}^d$ and $\boldsymbol{i}(t)\in \mathbb{R}^d$ at time $t$, respectively model the time-varying behavior and features.\\
\textbf{Interaction modeling layer.}
The interaction modeling layer updates the embedding of a user  and an item when the user interacts with the item. In particular, when an interaction $o_j = (u_j,i_j,t_j, \textbf{q}_j)$ is observed, the dynamic embedding of the involved user $u$ and item $i$ is updated.
For simplicity of notations we drop the $j$ subscript  in the following section to represent static embeddings as $\bar{\boldsymbol{u}}$ and $\bar{\boldsymbol{i}}$ and dynamic embedding as $\boldsymbol{u}(t)$ and $\boldsymbol{i}(t)$.\par
To obtain interaction-based embedding of $u$ and $i$, we consider their past interactions till time $t$ $\mathcal{O}^u(t) = \{o^u_1, o^u_2, \ldots o^u_p\} $ such that $t_p\leq t$  and  $\mathcal{O}^i(t) =\{o^i_1, o^i_2, \ldots o^i_q\} $ such that $t_q\leq t$, respectively.
  We use attention mechanism to compute the \textit{importance} of past interactions in determining the updated embedding of $u$ as:
\begin{equation}
    e^u_k(t) = a(\boldsymbol{W}^i\boldsymbol{i}_{k}(t_k^-), \boldsymbol{W}^u\boldsymbol{u}(t^-))+a(\boldsymbol{W}^q\boldsymbol{q}_k, \boldsymbol{W}^u\boldsymbol{u}(t^-))
\end{equation}

\noindent where $\boldsymbol{i}_k(t_k)$ represents the dynamic embedding of item occurring at $k$th interaction in $\mathcal{O}^u(t)$, $t^-$ represents the time right before the time $t$, $\boldsymbol{W}^u , \boldsymbol{W}^i \in \mathbb{R}^{d\times d}$, $\boldsymbol{W}^q \in \mathbb{R}^{d \times F}$  are the  weight matrices and $d$ and $F$ are the embedding size and the number of features associated with an interaction, respectively. The intuition is as follows, the first term computes importance of $i$'s features at the time of interaction to predict $u$'s future embedding. The second term introduces the level of contribution the interaction  features have towards the evolution of $u$. In our experiments, we used $a$ as the dot product between the two vectors. \par


Having computed the attention coefficients, $e^u(t)$, corresponding to all historical interactions involving $u$, we compute the new embedding of $u$ as:
\begin{align}
   \boldsymbol{u}(t) = &\sigma \bigg(\sum_{j, o_j \in \mathcal{O}^u(t)} \alpha_j^u(t)\boldsymbol{W}^i\boldsymbol{i}_{j} (t_j) \bigg), \alpha^u_{j}(t) =& \frac{exp(e^{u}_j(t))}{\sum_{k, o_k \in \mathcal{O}^u(t)} exp(e^{u}_k(t))},
\end{align}

\noindent  where $\sigma$ is introduced for non-linearity.
Here we have described an attention layer to update the embedding of user $u$. To update the embedding of $i$, we employ the same two operations with iteractions associated with the item. \\
\textbf{Influence modeling layer}
One of the major novelty of our method is that we introduce a time-varying self-attention based influence model for predicting user's future interest. The idea is to leverage the knowledge of evolution of a user's neighbors to  predict future embedding of the user. Modeling neighborhood influence in temporal interaction network poses specific challenge as the influence of an interaction on a user is driven by both the relation between users and time elapsed since the interaction.   \par
\begin{figure}[t]
       \includegraphics[width=\textwidth]{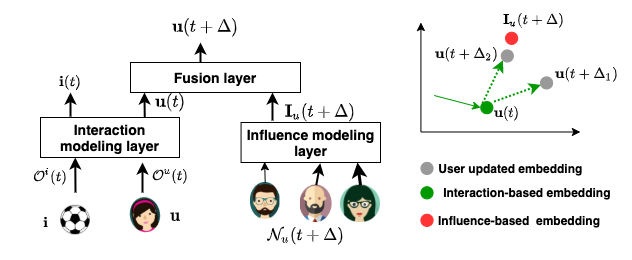}
     \caption{The IACN model: After an interaction $(u,i,t, \textbf{q})$, the dynamic embeddings of $u$ and
$i$ are updated in the Interaction modeling layer. The Influence modeling layer predicts the user embedding at time $t+\Delta$,  $\textbf{u}(t+\Delta)$ by taking influence vector $\textbf{I}_u(t+\Delta)$ into consideration. The figure on the  right side shows how influence modeling layer updates user embedding. As more time elapses, ($\Delta_2> \Delta_1$), the user embedding tends to be closer to $\textbf{I}_u(t)$.
 }
     \label{projection}

   \end{figure}
Our model  captures the influence of  $u$'s local neighborhood on $u$'s embedding by modeling a function that outputs a representation vector, influence embedding, $\boldsymbol{I}_u(t)$. This influence embedding is governed by an aggregation function parameterized by the temporal interaction sequence involving user neighborhood.  Influence-based embedding at time $t+\Delta$ is computed as:
\begin{equation}
    \boldsymbol{I}_u(t+\Delta) =  \sum_{v \in \mathcal{N}_u(t) t<t_v < t+\Delta} \theta_{v,u} exp(-\delta_u (t+\Delta-t_v))\boldsymbol{v}(t_v),
\end{equation}
\noindent where $\theta_{v,u}$ models the influence user $v$ has on $u$ and $exp(-\delta_u (t+\Delta-t_v))$  models decay of the influence over time with user-specific parameter $ \delta_u$ and $\mathcal{N}_u(t)$ is the local user neighborhood of $u$. To model the level of influence a user $v$ has on the other $u$, we again utilize the attention mechanism, i.e.,
\begin{equation}
    \theta_{v,u}= \begin{cases}
    a(\boldsymbol{W}_1^l\boldsymbol{v}(t_v),\boldsymbol{W}^l_2\boldsymbol{u}(t)) & \text{if } v \in \mathcal{N}_u(t),\\
    0 & \text{otherwise}
    \end{cases},
\end{equation}
\noindent where $\boldsymbol{W}_1^l$ and $\boldsymbol{W}^l_2$ are the weight parameters of the attention mechanism.  Due to peer engagement and affinity between users,  $\theta$ is sparse as users tend to indulge in discussions with users of their community. For validating this, we computed the average length of local user neighborhood in 'Wikipedia' dataset (described in section 5.1). We find that with 8227 users, the number of non-zero values in $\theta$ is 191,307. The average length of local user neighborhood is only 23.2.\\
\textbf{Fusion layer}
To integrate the signals from interaction layer and influence layer, we introduce a fusion layer. This layer predicts embeddings of user  at time $t$ by taking into account the user embedding, the influence embedding, and the time elapsed since  $u$'s last interaction, $\Delta$. The motivation behind constructing this layer is that a user interest keeps evolving even when it is not interacting with any item and as more time elapses the future embedding is farther from the user embedding. Furthermore, the interactions from the user local neighborhood influences the user interest which becomes more pronounced as more time elapses.
 To model this, we employ a kernel function such that the user embedding $\boldsymbol{u}(t+\Delta)$ will continue to deterministically decay (at
different rates for different users) from interaction-based embedding $\boldsymbol{u}(t)$ towards influence-based embedding $\boldsymbol{I}_u(t+\Delta)$. Thus, we extrapolate a user embedding at a future time as:

\begin{equation}
    {\boldsymbol{u}}(t+\Delta) =  \boldsymbol{u}(t)+ (\boldsymbol{I}_u(t+\Delta)-\boldsymbol{u}(t))   (1-exp(-\beta_u \Delta)),
\end{equation}
where $\beta_u$ is a parameter learned while training the model.
On the interval $[t, t+\Delta)$, the $u$'s embedding follows an exponential curve that begins at $\boldsymbol{u}(t)$, when $\Delta \rightarrow 0$ and decays towards $\boldsymbol{I}_u(t)$ ( as $t \rightarrow \infty$, if extrapolated). \\
\textbf{Recommendation layer.}
Once we predict users' embeddings at time $t+\Delta$, we predict the embedding of the next item. For this we use the updated user embedding
$\boldsymbol{u}(t + \Delta)$ and the  embedding of item that $u$ last interacted with at time $t$,  $\boldsymbol{i}(t )$.  The predicted item embedding is:
\begin{equation}
    \boldsymbol{\hat{i}}(t + \Delta) = \boldsymbol{W}[ \boldsymbol{u}(t + \Delta), \bar{\boldsymbol{u}}, \boldsymbol{i}(t ) , \bar{\boldsymbol{i}} ] +\boldsymbol{B}  ,
\end{equation}

\noindent where $\boldsymbol{W}$ is the weight matrix and $\boldsymbol{B}$ bias vector which make the linear layer. Then we recommend the items with the closest embedding with the predicted  embedding. This step can be done in near-constant time by using LSH ~\cite{gionis1999similarity}.
\vspace{-4mm}
\subsection{Network training}
\vspace{-2mm}
We train our model to minimize the Euclidean distance between the predicted item embedding and the actual item embedding everytime a user interacts with an item. We calculate the total loss as,
\begin{align*}
 \begin{split}
     \mathcal{L} = {}&\sum_{(u,i,t,\boldsymbol{q}) \in \mathcal{O}} ||\boldsymbol{\hat{i}}(t)-[\boldsymbol{\bar{i}}, \boldsymbol{i}(t)]||_2 + \lambda_U ||\boldsymbol{u}(t)-\boldsymbol{u}(t^-)||_2 + \lambda_I ||\boldsymbol{i}(t) - \boldsymbol{i}(t^-)||_2,
 \end{split}
\end{align*}

\noindent where $\lambda_U$ and $\lambda_I$ are regularization parameters for temporal smoothness of user and item embeddings, respectively.

\vspace{-5mm}
\section{Experimental Settings}
\vspace{-3mm}
To comprehensively evaluate the performance of our proposed
IACN  model, we design different strategies to evaluate the effectiveness of the model.  Our experiments are designed to answer the following research questions:
\begin{enumerate}
    \item  \textbf{RQ1:} How does IACN perform compared  with other state-of-the-art recommendation models?
\item \textbf{RQ2:} What is the influence of various components in the IACN architecture?

\end{enumerate}
\textbf{Datasets.}
We used 4 public datasets and followed the same preprocessing steps as used in ~\cite{kumar2019predicting}. Thus,we selected $1000$ most active items in each dataset.
\vspace{-2mm}
\begin{itemize}

\item \textbf{Wikipedia dataset:} Public dataset consisting of one month of edits made on Wikipedia pages
 \footnote{\url{https://meta.wikimedia.org/wiki/Data\textunderscore dumps}.}
obtained from ~\cite{kumar2019predicting}. This dataset contains $1000$ items, $10,000$ most active users, resulting in $672,447$ interactions.

    \item \textbf{Reddit post dataset:} We processed reddit \footnote{ http://files.pushshift.io/reddit/} forum dataset, which consists of one month of posts made by users. We first samples $1000$ most active reddit post and the users who made at least $5$ posts on the selected posts. This resulted in $13,840$ users and a total of $121,258$ interactions.

\item \textbf{Yelp review dataset:} We obtained this dataset from the yelp dataset challenge\footnote{https://www.kaggle.com/yelp-dataset/yelp-dataset}. We first selected top $1000$ businesses with most number of reviews and users who made at least $5$ reviews on the selected businesses. This resulted in $5325$ users and $110,839$ interactions.
\item \textbf{StackOverFlow dataset:} We also gathered data from the popular question-answering website, StackOverFlow\footnote{https://archive.org/details/stackexchange}.  For this dataset also, we extracted users who made at least $5$ posts. There are $4,125$ users and $20,719$ posts in this dataset.
\end{itemize}
\vspace{-2mm}
These datasets, in addition to varying in size of users and density of interactions, also comprise of different users' behavior in terms of repetitive item consumption.
In Wikipedia, Reddit, and StackOverFlow a user interacts with
the same item consecutively in $79\%$, $77\%$  and $62\%$ interactions, respectively, while in Yelp it occurs only in $0.004\%$ of interactions. Naturally, we expect that the Yelp dataset is the most challenging one to model.\\
Code available at \url{https://github.com/shalini1194/IACN}.\\
\textbf{Metrics.}
We evaluate forum recommendation performance using the  mean reciprocal rank (MRR)
and recall@10.
MRR is a standard ranking metric formulated as: $MRR = \frac{1}{\text{rank}_{\text{pos}}}$, where $\text{rank}_{\text{pos}}$ denotes the rank of positive item.
Recall@10 is the fraction of  ground truth items ranked in the top $10$ recommended items.\\
\textbf{Comparison Approaches.} To verify the performance gain of IACN, we compare its performance with various state-of-the-art models which can be categorized into four classes:
\begin{enumerate}
    \item RNN based models: This category comprises of RNN based models such as LSTM ~\cite{hidasi2015session},  RRN ~\cite{wu2017recurrent} among others. RNN uses only static embeddings to represent items and predicts users' embedding based on the items they have interacted with. RRN is widely used method and generates dynamic user and item embeddings based on the item and user interaction sequence independently. Both these models take one-hot vector of items as inputs.
    \item Co-evolutionary models: These models update both user and item embedding when a user interacts with an item. We compare our model with JODIE ~\cite{kumar2019predicting} and Deep Co-evolve ~\cite{dai2016deep}.
    Both the models use RNN to learn representations of users and items. Deep-Coevolve uses the point process technique to predict the intensity of interaction between user and item, while JODIE  uses Euclidean distance between the learned representation to predict the next item to recommend.
    \item Temporal Network Embedding: Temporal Network Embedding models are used to generate embedding of nodes of a temporal network. HTNE ~\cite{zuo2018embedding} is a state-of-the-art model for temporal network embedding which integrates the Hawkes process into
network embedding so as to capture the influence of historical
neighbors on the current neighbors

    \item Social Network:  We compare our method with GraphRec ~\cite{fan2019graph}  that combines the information from social network and interaction network to predict user embedding. However, it does not consider the temporal nature of the setting.
\end{enumerate}
\begin{table*}[t]

\centering
\caption{ Performance comparison on four datasets for all methods. The best and the second best results are highlighted by
\textbf{boldface} and \underline{underlined } respectively. Gain$\%$ denotes the performance improvement of IACN over the best baseline.}
\begin{tabular}{|c|rr|rr|rr|rr|}
\toprule
\multicolumn{1}{|c|}{Methods}& \multicolumn{2}{c|}{Wikipedia}                           & \multicolumn{2}{c|}{Reddit}                              & \multicolumn{2}{c|}{Yelp}                                & \multicolumn{2}{c|}{StackOverFlow}                           \\
\multicolumn{1}{|l|}{}                  & \multicolumn{1}{c}{MRR} & \multicolumn{1}{c|}{Recall@10} & \multicolumn{1}{c}{MRR} & \multicolumn{1}{c|}{Recall@10} & \multicolumn{1}{c}{MRR} & \multicolumn{1}{c|}{Recall@10} & \multicolumn{1}{c}{MRR} & \multicolumn{1}{c|}{Recall@10} \\
\hline

LSTM  ~\cite{hidasi2015session}         & 0.329 & 0.455 & 0.205  & 0.251 & 0.007 & 0.009  & 0.014 & 0.017  \\
\hline
RRN  ~\cite{wu2017recurrent}          & 0.522 & 0.617 & 0.290  & 0.312 & 0.013 & 0.020  & 0.019 & 0.019  \\
\hline
HTNE ~\cite{zuo2018embedding}          & 0.500 & 0.624 & 0.211  & 0.313 & 0.012 & 0.014  & \underline{0.100} & \underline{0.178}  \\
\hline
GrapRec ~\cite{fan2019graph}            & 0.634 & \underline{0.823} & 0.621  & 0.815 & 0.006 & 0.009  & 0.012 & 0.041  \\
\hline
DeepCo-evolve ~\cite{dai2016deep} & 0.515 & 0.563 & 0.271  & 0.405 & 0.006 & 0.008  & 0.017 & 0.019  \\
\hline
JODIE   ~\cite{kumar2019predicting}       & \underline{0.746} & 0.822 & \underline{0.755}  & \underline{0.919} & \underline{0.014} & \underline{0.020}  & 0.058 & 0.063  \\
\hline
IACN           & \textbf{0.796} & \textbf{0.861} & \textbf{0.869}  & \textbf{0.922} & \textbf{0.015} & \textbf{0.026}  & \textbf{0.106} & \textbf{0.280}  \\
\hline
Gain \%        & 6.702 & 4.617 & 15.099 & 0.326 & 7.143 & 30.000 & 6.000 & 57.303\\
\bottomrule

\end{tabular}
\label{performance}

\end{table*}

\vspace{-6mm}
\subsection{Performance Comparison (\textbf{RQ1})}
\vspace{-3mm}
Table ~\ref{performance} compares the performance of IACN with the six
state-of-the-art methods.  We make the following
observations from the results.
 IACN significantly outperforms all baselines in all datasets across both the metrics.
 GraphRec performs better than HTNE for Reddit and Wikipedia dataset. We believe that one of the reasons is the high volume of interactions in less timespan for these datasets. Due to this, the effect of time intervals between interactions is not observed here. HTNE models the impact of time intervals between interactions, which results in its better performance for Yelp and StackOverFlow compared to GraphRec.
We find that for StackOverFlow dataset HTNE performs better than JODIE. This can be attributed to the idea that user-user affinity is more pronounced due to peer-engagement and depth of discussion on these platforms ~\cite{paranjape2017motifs}. The fact that IACN outperforms co-evolutionary models  confirms our hypothesis that it is important to consider both influence-based and interaction-based signals to predict embedding of user. \par
\begin{table*}[t]
\caption{Ablation analysis  on four datasets.}
\label{ablation}
\begin{tabular}{|c|rr|rr|rr|rr|}

\toprule
Methods                            & \multicolumn{2}{c|}{Wikipedia}                           & \multicolumn{2}{c|}{Reddit}        & \multicolumn{2}{c|}{Yelp}                                & \multicolumn{2}{c|}{StackOverFlow}                       \\

                                   & \multicolumn{1}{c}{MRR} & \multicolumn{1}{c|}{Recall@10} & \multicolumn{1}{c}{MRR} & \multicolumn{1}{c|}{Recall@10} & \multicolumn{1}{c}{MRR} & \multicolumn{1}{c|}{Recall@10} & \multicolumn{1}{c}{MRR} & \multicolumn{1}{c|}{Recall@10} \\
                                   \hline
IACN - Influence   & 0.776                   & 0.833                         & 0.717                  & 0.919                         & 0.009                   & 0.014                         & 0.050                   & 0.059                         \\
\hline
IACN-Attention+RNN & 0.786                   & 0.848                         & 0.717                  & 0.920                         & 0.008                   & 0.011                         & 0.056                   & 0.059                         \\
\hline
IACN-Fusion+LatentCross &0.612 & 0.776& 0.702&0.918 &0.011 &0.018 &0.072 &0.012\\
\hline
IACN                               & 0.796                   & 0.861                         & 0.869                   & 0.922                        & 0.015                   & 0.026                         & 0.106                   & 0.280  \\
\bottomrule

\end{tabular}

\end{table*}
\vspace{-5mm}
\subsection{Analysis of IACN (\textbf{RQ2})}
\vspace{-3mm}
Table ~\ref{ablation} shows the performance comparison of variation of IACN. We describe the variants and discuss the result drop caused by them:\\
\textbf{IACN-Influence:} Removing the influence modeling layer results in a co-evolutionary model with attention mechanism to update the embedding. We find that removing the influence modeling layer results in drop of IACN performance, revealing that it is useful to model the influence of other users on user interest evolution. \\
\textbf{IACN-Attention+RNN:} In this variant, we replace the attention in the interaction modeling layer with RNN. The drop in performance indicates that attention mechanism is better able to predict the  embedding of user and item by adaptively assigning weights to the past interactions.\\
\textbf{IACN-Fusion+LatentCross} In this variant of IACN, we replace our Fusion layer with LatentCross ~\cite{beutel2018latent}. Essentially, we take an element-wise product of user embedding $u(t)$ and the time context vector, $\boldsymbol{w}_t = \boldsymbol{w}*\Delta$, where, $\boldsymbol{w}$ is initialized by $0$-mean Gaussian function and $\Delta$ is the elapsed time since user's last interaction. Then, we add the influence-based embedding to the resultant vector.
\vspace{-1mm}
\begin{equation*}
    \boldsymbol{u}(t+\Delta) = (\boldsymbol{1}+\boldsymbol{w}_t)*\boldsymbol{u}(t)+\boldsymbol{I}_u(t+\Delta)
\end{equation*}

Using LatentCross instead of our fusion layer degrades performance of IACN showing that fusions layer is better then LatentCross.

\section{Conclusion and Future Work}
\vspace{-3mm}
In this paper, we proposed a novel model to predict dynamic embedding of user and item which takes into account both reasons of evolution of user interest, namely, interaction with an item and influence from other users. IACN utilizes attention mechanism to update  embedding of users and items when they interact. It also models the influence of activities by local user neighborhood on the user interest. For future work, instead of modeling the relation between each pair of users, one can model the group the users and use the embedding of local user group to predict the evolution of local neighborhood of user.
\bibliographystyle{splncs04}
\bibliography{ref.bib}

\begin{thebibliography}{10}
\providecommand{\url}[1]{\texttt{#1}}
\providecommand{\urlprefix}{URL }
\providecommand{\doi}[1]{https://doi.org/#1}

\bibitem{beutel2018latent}
Beutel, A., Covington, P., Jain, S., Xu, C., Li, J., Gatto, V., Chi, E.H.:
  Latent cross: Making use of context in recurrent recommender systems. In:
  Proceedings of the Eleventh ACM International Conference on Web Search and
  Data Mining. pp. 46--54. ACM (2018)

\bibitem{chen2019social}
Chen, C., Zhang, M., Liu, Y., Ma, S.: Social attentional memory network:
  Modeling aspect-and friend-level differences in recommendation. In:
  Proceedings of the Twelfth ACM International Conference on Web Search and
  Data Mining. pp. 177--185. ACM (2019)

\bibitem{dai2016deep}
Dai, H., Wang, Y., Trivedi, R., Song, L.: Deep coevolutionary network:
  Embedding user and item features for recommendation. arXiv preprint
  arXiv:1609.03675  (2016)

\bibitem{fan2019graph}
Fan, W., Ma, Y., Li, Q., He, Y., Zhao, E., Tang, J., Yin, D.: Graph neural
  networks for social recommendation. In: The World Wide Web Conference. pp.
  417--426. ACM (2019)

\bibitem{gionis1999similarity}
Gionis, A., et~al.: Similarity search in high dimensions via hashing

\bibitem{hidasi2015session}
Hidasi, B., Karatzoglou, A., Baltrunas, L., Tikk, D.: Session-based
  recommendations with recurrent neural networks. arXiv preprint
  arXiv:1511.06939  (2015)

\bibitem{iwata2013discovering}
Iwata, T., Shah, A., Ghahramani, Z.: Discovering latent influence in online
  social activities via shared cascade poisson processes. In: Proceedings of
  the 19th ACM SIGKDD international conference on Knowledge discovery and data
  mining. pp. 266--274. ACM (2013)

\bibitem{kumar2019predicting}
Kumar, S., Zhang, X., Leskovec, J.: Predicting dynamic embedding trajectory in
  temporal interaction networks. In: Proceedings of the 25th ACM SIGKDD
  International Conference on Knowledge Discovery \& Data Mining. pp.
  1269--1278 (2019)

\bibitem{lu2018coevolutionary}
Lu, Y., Dong, R., Smyth, B.: Coevolutionary recommendation model: Mutual
  learning between ratings and reviews. In: Proceedings of the 2018 World Wide
  Web Conference. pp. 773--782 (2018)

\bibitem{nguyen2018continuous}
Nguyen, G.H., Lee, J.B., Rossi, R.A., Ahmed, N.K., Koh, E., Kim, S.:
  Continuous-time dynamic network embeddings. In: Companion Proceedings of the
  The Web Conference 2018. pp. 969--976 (2018)

\bibitem{paranjape2017motifs}
Paranjape, A., Benson, A.R., Leskovec, J.: Motifs in temporal networks. In:
  Proceedings of the Tenth ACM International Conference on Web Search and Data
  Mining. pp. 601--610 (2017)

\bibitem{song2019session}
Song, W., Xiao, Z., Wang, Y., Charlin, L., Zhang, M., Tang, J.: Session-based
  social recommendation via dynamic graph attention networks. In: Proceedings
  of the Twelfth ACM International Conference on Web Search and Data Mining.
  pp. 555--563. ACM (2019)

\bibitem{vaswani2017attention}
Vaswani, A., Shazeer, N., Parmar, N., Uszkoreit, J., Jones, L., Gomez, A.N.,
  Kaiser, {\L}., Polosukhin, I.: Attention is all you need. In: Advances in
  neural information processing systems. pp. 5998--6008 (2017)

\bibitem{wu2017recurrent}
Wu, C.Y., Ahmed, A., Beutel, A., Smola, A.J., Jing, H.: Recurrent recommender
  networks. In: Proceedings of the tenth ACM international conference on web
  search and data mining. pp. 495--503. ACM (2017)

\bibitem{xia2009ballot}
Xia, M., Huang, Y., Duan, W., Whinston, A.B.: Ballot box communication in
  online communities. Communications of the ACM  \textbf{52}(9),  138--142
  (2009)

\bibitem{zuo2018embedding}
Zuo, Y., Liu, G., Lin, H., Guo, J., Hu, X., Wu, J.: Embedding temporal network
  via neighborhood formation. In: Proceedings of the 24th ACM SIGKDD
  International Conference on Knowledge Discovery \& Data Mining. pp.
  2857--2866. ACM (2018)

\end{thebibliography}
\end{document}